\documentclass{ws-procs975x65}

\begin{document}

\title{EFFECTS OF THE SINGULAR SELF-FIELD ON THE MOTION OF AN EXTENDED BODY}

\author{ABRAHAM I. HARTE}

\address{Enrico Fermi Institute, University of Chicago\\
Chicago, IL 60637 U.S.A.\\
E-mail: harte@uchicago.edu}

\begin{abstract}
A formalism is described that greatly simplifies the derivation of scalar, electromagnetic, and gravitational self-forces and self-torques acting on extended bodies in curved spacetimes. Commonly-studied aspects of these effects are normally dominated by the so-called ``regular" component of a body's self-field. The only consequence of the remaining (much larger) portion of the self-field turns out to be very simple. It exerts forces and torques that effectively renormalize all multipole moments of the body's stress-energy tensor in its laws of motion.
\end{abstract}

\bodymatter

\section{Introduction}

Problems related to the motion of compact bodies in curved spacetimes have been studied in various contexts for nearly a century. At least three broad (and not entirely independent) regimes are commonly considered. The most completely developed of these is concerned with the behavior of extended test bodies. A formalism developed by Dixon \cite{Dix74} elegantly solves this problem -- among others -- in essentially arbitrary background spacetimes. The body in question may be highly relativistic and nonpsherical. If its self-field can be ignored, multipole approximations for the laws of motion are known to all orders for both gravitational and electromagnetic interactions. Another regime of interest is covered by the post-Newtonian approximations \cite{PNReview}. These allow masses to interact with each other, and even to radiate, but are usually restricted to low speeds and small shear stresses. Lastly, it is common to consider the motion of compact relativistic objects with a small, but non-negligible amount of self-interaction \cite{PoissonRev}. Such systems were first studied in electromagnetism in flat spacetime, although they arise in almost any theory with matter coupled to a long-range field. In general relativity, the typical application of the self-force problem is to a stellar-sized object spiraling into a supermassive black hole in the core of a galaxy.

This final regime is the one of interest here. At least in a non-relativistic limit, It has long been known that the acceleration $\vec{a}$ of a small electric charge $q$ with mass $m$ in the presence of an external electric field $\vec{E}_\mathrm{ext}$ is approximately given by
\begin{equation}
m \vec{a} \simeq q \vec{E}_\mathrm{ext} - \delta m \vec{a} + \frac{2}{3} q^2 \frac{\mathrm{d} \vec{a}}{\mathrm{d} t},
\label{ALD}
\end{equation}
for some $\delta m$ depending on the details of the charge's internal structure. The first term on the right-hand side is the ordinary Lorentz force experienced by a small test charge. The remaining two terms are due to the self-field. One acts to add an effective inertia $\delta m$ to the system. This arises from the energy associated with the charge's self-field. The last term in \eqref{ALD} is due to radiation reaction.

A very similar split arises in relativistic electrodynamics, as well as for matter coupled to a Klein-Gordon field or to linearized general relativity. This is also true if the background spacetime is strongly curved. In each case, the effective shift in the body's inertia is not often interesting. It also diverges if one tries to take a point particle limit in the usual sense. This has led to methods intended to extract only the last term in \eqref{ALD}. One of most useful of these observations is due to Detweiler and Whiting \cite{DetWhiting}. Schematically, they noticed that all known self-force calculations performed up to that point could be put into the form
\begin{equation}
m_\mathrm{eff} \vec{a} \simeq q \vec{E}_\mathrm{eff},
\label{SchematicDW}
\end{equation}
where $\vec{E}_\mathrm{eff}$ was a particular solution of the homogeneous field equations near the particle. This was extremely useful, as $\vec{E}_\mathrm{eff}$ usually varied slowly near the object, and was very close to the external field. All effects of the rapidly-varying ``singular self-field'' $\vec{E}_\mathrm{S} \equiv \vec{E} - \vec{E}_{\mathrm{eff}}$ (so named due to its behavior for point particles) were effectively absorbed into the renormalized mass $m_\mathrm{eff}$. The object behaves as though it were a test particle moving in the field $\vec{E}_\mathrm{eff}$.

\section{A general self-force formalism}

Results like these have now been rigorously derived for bodies coupled to Klein-Gordon \cite{HarteScalar} and Maxwell \cite{HarteEMNew} fields in arbitrary background spacetimes, as well as for those coupled to the linearized Einstein equations. This is done by modifying Dixon's extended-body formalism. One first introduces linear and angular momenta for an extended body by defining a ``time''-dependent map $\mathcal{P}_\xi(s)$ from a 10-dimensional space of generalized Killing fields (GKFs) into $\mathbb{R}$. This involves the singular self-field. Consider the scalar self-force problem for simplicity. All components of momentum associated with a body having stress-energy tensor $T^{ab}$ and scalar charge density $\rho$ in a spacetime with fixed metric $g_{ab}$ evolve according to
\begin{equation}
	\frac{\mathrm{d}}{\mathrm{d} s} \mathcal{P}_\xi = \int_\Sigma \mathrm{d}S \left[ \frac{1}{2} T^{ab} \mathcal{L}_\xi g_{ab} + \rho \mathcal{L}_\xi 			\phi_\mathrm{eff} + \frac{1}{2} \int \mathrm{d} V' \rho \rho' \mathcal{L}_\xi G_\mathrm{S} (x,x') \right].
	\label{PDot}
\end{equation}
The first term here is responsible for purely gravitational forces and torques. This usually couples to the quadrupole and higher moments of the body's stress-energy tensor, and is present even for test bodies.  The second term denotes the ordinary force on a scalar charge placed in the effective field $\phi_\mathrm{eff} = \phi-\phi_{\mathrm{S}}$. As in \eqref{SchematicDW}, $\phi_{\mathrm{eff}}$ satisfies the homogeneous field equations. In many cases, the force or torque that it exerts can therefore be adequately approximated by the first term in a multipole expansion. Lastly, the symmetric two-point scalar $G_\mathrm{S}(x,x')$ is the Green function used to define the singular self-field $\phi_\mathrm{S}$.

In flat or de Sitter spacetime, all of the GKFs are genuine Killing fields. Eq. \eqref{PDot} therefore reduces to
\begin{equation}
	\frac{\mathrm{d}}{\mathrm{d} s} \mathcal{P}_\xi = \int_\Sigma \mathrm{d}S \rho \mathcal{L}_\xi \phi_{\mathrm{eff}} .
\end{equation}
This is an exact result. All remnants of the ``singular'' self-field -- which is actually bounded for any smooth charge distribution --  have been absorbed into the definition of the momenta. The resulting quantities clearly evolve as though the object were a test body placed in the homogeneous field $\phi_{\mathrm{eff}}$. This is accomplished non-perturbatively without any detailed knowledge of solutions to the field equations.

It is interesting to ask what happens in curved spacetimes that are not maximally symmetric. In these cases, the last term in \eqref{PDot} is normally very small, but does not vanish entirely. Its interpretation turns out to be very simple. We have already noted that the singular field is included in the linear and angular momenta, and may therefore be thought of as renormalizing the na\"{\i}ve definitions of these quantities. While the analogy is not perfect, one may loosely interpret this effect as being due to the energy contained in the ``bound'' portion of the body's self-field. Of course, that energy is not localized to one point. Its distribution may be described in terms of multipole moments. The first two of these essentially shift the ``bare'' linear and angular momenta. One may, of course, define quadrupole and higher moments of the energy density as well. These couple to the laws of motion only in the presence of a curved background. It is exactly this effect that is represented by the term involving $\mathcal{L}_\xi G_\mathrm{S}$ in \eqref{PDot}. If the metric varies smoothly near the body, all multipole moments of $T^{ab}$ are effectively renormalized by this term \cite{HarteSingField}. To quadrupole order, one has
\begin{equation}
	\frac{\mathrm{d}}{\mathrm{d} s} \mathcal{P}_\xi = - \frac{1}{6} (J^{abcd} + \delta J^{abcd} ) \mathcal{L}_\xi R_{abcd} + \int_\Sigma \mathrm{d}S \rho \mathcal{L}_\xi \phi_{\mathrm{eff}} .
\end{equation}
$J^{abcd}$ is the quadrupole moment of $T^{ab}$. It comes from the first term on the right-hand side of \eqref{PDot}. $\delta J^{abcd}$, by contrast, arises from the last term in that equation. One can write it entirely in terms of a trace-free two index ``mass quadrupole'' in the case that $R_{ab} = \lambda g_{ab}$ for some (possibly vanishing) constant $\lambda$.

The basic point of the Detweiler-Whiting idea has therefore been derived in a very general context. A scalar charge in a fixed background spacetime evolves as though it were an extended test body moving in metric $g_{ab}$ and scalar field $\phi_\mathrm{eff}$. The various parameters associated with that body -- its mass, angular momentum, mass quadrupole, etc. -- all depend on the singular field, but this is often unimportant for an external observer. All complications due to the rapidly-varying $\phi_\mathrm{S}$ are effectively hidden. The same result holds in Maxwell theory and general relativity linearized off an arbitrary background.


\begin{thebibliography}{9}
\bibitem{Dix74} W. G. Dixon, \textit{Phil. Trans. R. Soc. A} \textbf{277}, 59 (1974)
\bibitem{PNReview} T. Futamase and Y. Itoh, \textit{Living Rev. Rel.} \textbf{10}, 2 (2007)
\bibitem{PoissonRev} E. Poisson, \textit{Living Rev. Rel.} \textbf{7}, 6 (2004)
\bibitem{DetWhiting} S. Detweiler and B. F. Whiting, \textit{Phys. Rev. D} \textbf{67}, 024025 (2003)
\bibitem{HarteScalar} A. I. Harte, \textit{Class. Quant. Grav.} \textbf{25}, 235020 (2008)
\bibitem{HarteEMNew} A. I. Harte, \textit{Class. Quant. Grav.} \textbf{26}, 155015 (2009)
\bibitem{HarteSingField} A. I. Harte, arXiv: 0910.4614
\end{thebibliography}

\end{document}